\documentclass[aps,prl,twocolumn,superscriptaddress]{revtex4}

\usepackage{graphicx,psfrag}
\usepackage[latin1]{inputenc}
\usepackage{natbib}
\usepackage{amsmath, amsthm, amssymb}

\newcommand{\ds}{\displaystyle}
\newcommand{\mb}{\mathbf}
\newcommand{\notes}[1]{}

\newcommand{\beq}{\begin{equation}}
\newcommand{\eeq}{\end{equation}}
\newcommand{\beqnn}{\begin{equation*}}
\newcommand{\eeqnn}{\end{equation*}}
\newcommand{\beqas}{\begin{eqnarray*}}
\newcommand{\eeqas}{\end{eqnarray*}}
\newcommand{\beqa}{\begin{eqnarray}}
\newcommand{\eeqa}{\end{eqnarray}}
\newcommand{\p}{\partial}

\begin{document}

\title{Partial spin reversal in magnetic deflagration}

\author{S. V\'elez}
\altaffiliation{present affiliation: CIC nanoGUNE, 20018 Donostia-San Sebastian, Basque Country, Spain}
\affiliation{Grup de Magnetisme, Dept. de F\'isica, Universitat de Barcelona, Spain}
\affiliation{Department of Physics, New York University, New York, New York 10003, USA}
\author{P. Subedi}
\affiliation{Department of Physics, New York University, New York, New York 10003, USA}
\author{F. Maci\`a}
\affiliation{Grup de Magnetisme, Dept. de F\'isica, Universitat de Barcelona, Spain}
\affiliation{Department of Physics, New York University, New York, New York 10003, USA}
\author{S. Li}
\affiliation{Department of Physics, City College of New York, CUNY, New York, New York
10031, USA}
\author{M. P. Sarachik}
\affiliation{Department of Physics, City College of New York, CUNY, New York, New York
10031, USA}
\author{\\ J. Tejada}
\affiliation{Grup de Magnetisme, Dept. de F\'isica, Universitat de Barcelona, Spain}
\author{S. Mukherjee}
\author{G. Christou}
\affiliation{Department of Chemistry, University of Florida, Gainesville, Florida 32611,
USA}
\author{A. D. Kent}
\affiliation{Department of Physics, New York University, New York, New York 10003, USA}
\date{\today}

\begin{abstract}
The reversal of spins in a magnetic material as they relax toward equilibrium is accompanied by the release of Zeeman energy which can lead to accelerated spin relaxation and the formation of a well-defined self-sustained propagating spin-reversal front known as magnetic deflagration. To date, studies of Mn$_{12}$-acetate single crystals have focused mainly on deflagration in large longitudinal magnetic fields and found a fully spin-reversed final state.  We report a systematic study of the effect of transverse magnetic field on magnetic deflagration and demonstrate that in small longitudinal fields the final state consists of only partially reversed spins. Further, we measured the front speed as a function of applied magnetic field. The theory of magnetic deflagration, together with a modification that takes into account the partial spin reversal, fits the transverse field dependence of the front speed but not its dependence on longitudinal field. The most significant result of this study is the finding of a partially spin-reversed final state, which is evidence that the spins at the deflagration front are also only partially reversed.
\end{abstract}

\maketitle

\section{Introduction}
Spin relaxation in a magnetic system in a magnetic field can release heat and lead to a thermally driven instability with a well defined self-sustained traveling spin-reversal front known as magnetic deflagration.  This is analogous to chemical combustion, a reaction-diffusion process in which energy is released locally and diffuses to neighboring sites, ultimately spreading throughout the material, like a forest fire.  A deflagration front develops when the rate of local energy released exceeds the rate of diffusion of energy away from the local site.

Magnetic deflagration has been studied in single crystals of the molecular magnet Mn$_{12}$-acetate, [Mn$_{12}$O$_{12}$(O$_{2}$CCH$_{3}$)$_{16}$(H$_{2}$O)$_{4}$]$ \cdot $2CH$_{3}$CO$_{2}$H$\cdot $ 4H$_{2}$O (hereafter, denoted Mn$_{12}$-ac) \cite{suzuki,quantumdeflagration,maciaepl2006,McHugh2007,macia2009,McHugh2009,review}, as well as in other systems, including manganites \cite{macia2007} and intermetallic components \cite{velez2010}.  Unlike its chemical analogue, magnetic deflagration is reversible and non-destructive, allowing repeated measurements on a given sample for a broad range of parameters.  Moreover, Subedi {\it et al.} \cite{subedi2013} have shown that the onset of deflagration as well as the speed of propagation of the spin reversal front can be controlled by externally applied magnetic fields in Mn$_{12}$-ac, allowing in-depth investigations that are relevant to deflagration in other contexts.

To date, studies of Mn$_{12}$-ac single crystals have focused mainly on deflagration in large longitudinal magnetic fields and found a fully spin-reversed final state. Here we report systematic studies of deflagration in Mn$_{12}$-ac in the presence of large transverse magnetic fields, where the process can result in incomplete spin reversal that leaves the crystal in a partially magnetized, blocked, final state determined by the magnitude and direction of the externally applied magnetic field. These results are evidence that the spin reversal is only partial at the deflagration front and thus important in understanding deflagration processes even in larger longitudinal fields, that is, fields at which the final state measured after the process are fully spin-reversed.
\section{Background}

Mn$_{12}$-ac molecules have a core of twelve Mn atoms that are exchange-coupled through oxygen bridges to yield a net total spin $S=10$ at low temperature; the Mn$_{12}$-ac molecules can be modeled as a single giant spin with no internal degrees of freedom.  A large uniaxial magnetic anisotropy leads to magnetic bistability at low temperature. In the presence of an applied field $\mathbf{H}=(H_x, H_z)$ the simplest spin Hamiltonian  takes the form:
\begin{equation}
\mathcal{H}= -DS_z^2-g\mu_B H_x S_x - g\mu_B H_z S_z.
\label{hamintro4}
\end{equation}
The first term represents the uniaxial anisotropy and the second and third terms are the Zeeman energy corresponding to the field applied perpendicular (transverse) and parallel (longitudinal) to the uniaxial easy axis direction, respectively.

\begin{figure}[t]
\includegraphics[width=\columnwidth]{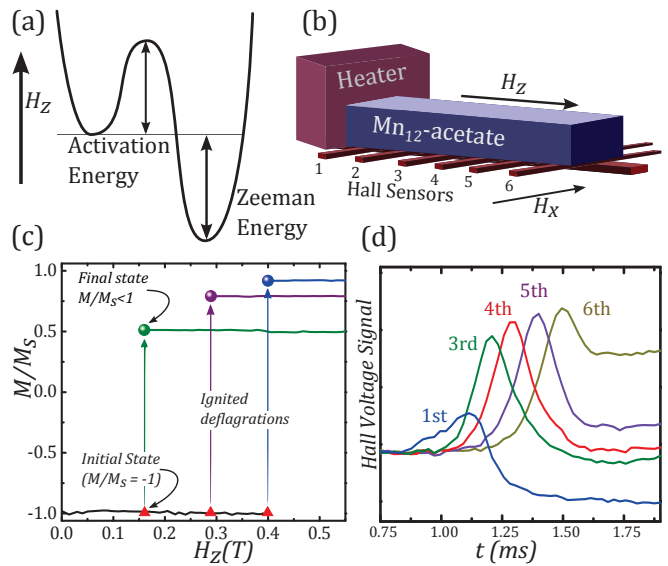}
\caption{\small{ (a) Double well potential energy showing activation and Zeeman energies of Mn$_{12}$-ac. (b) Schematic of the single crystal Mn$_{12}$-ac, Hall sensors, heater, and the directions of the applied magnetic fields.  (c) Initially the sample is saturated such that $M/M_s=-1$. Then at a set longitudinal and transverse field a heat pulse is applied (triangles). The final magnetization state is indicated by the solid circles. The curves shown are at $H_x=0$~T. (d) Signals from the Hall sensor array as a function of time for at $H_x=1.5$ T and $H_z=0.65$ T.}}
\label{fig1}
\end{figure}

The result is an effective double-well potential shown in Fig.~\ref{fig1}(a) with a barrier separating spin-up and spin-down projections. The activation energy for spin-reversal in the absence of a magnetic field is the full anisotropy barrier $U=DS^2 \approx 65$ K. As shown in Fig.~\ref{fig1}(a), a longitudinal magnetic field $H_z$ tilts the potential, reducing the activation energy and increasing the relaxation rate.  Quantum tunneling of the magnetization occurs at specific resonant values of the longitudinal magnetic field ($H_z=kD/(g\mu_B) \simeq  0.45k$ T, where $k$ is an integer), corresponding to applied fields at which levels with opposite spin-projections on the easy axis have avoided energy level crossings.  At the same time, increasing the longitudinal magnetic field increases the Zeeman energy released when spins reverse, $\Delta E = 2g \mu_BH_zS$.  A longitudinal field thus changes both the activation energy and the energy released into the system (but not independently).

The effect of a transverse field $H_x$ is to mix the eigenstates of $S_z$ and thus states with opposite spin-projections, which reduces the activation energy.  Unlike the longitudinal field, a transverse field does not change the energy released to lowest order. However, a transverse field greatly enhances the tunneling rates, which increases the magnetic relaxation rates, and enables deflagration at small longitudinal bias fields, the conditions we explore in this paper.

Using a trigger pulse to initiate spin-reversal, measurements in Mn$_{12}$-ac by Subedi {\it et al.} \cite{subedi2013} identified a sharp crossover between two distinct regimes. (1) At low bias and small fields, heat spreads throughout the crystal relatively slowly and slows down as it travels; in this thermal regime, the energy spread is guided by diffusion of the input pulse energy.  (2) For high bias and/or transverse fields, a self-sustained spin-reversal front is found, the front is driven by the Zeeman energy, and the speed of propagation is much higher and constant; this is referred to as the deflagration regime.

\section{Experimental Procedure}
A one dimensional array of Hall sensors (active area 20 $\times$ 100 $\mu$m$^2$ with 200 $\mu$m separation) was used to measure the magnetization of Mn$_{12}$-ac crystals at different positions (see Ref.~\cite{subedi2013} for details). Three different Mn$_{12}$-ac samples were studied and similar results were obtained in each case. Here we present representative data from one sample with dimensions $0.4\times0.4\times1.6$ mm$^3$. The samples were placed on the Hall sensor array as schematically shown in Fig. 1(b). A thin film heater ($R\approx1.32$ k$\Omega$ at 0.4 K) was mounted on one end of the crystal. A six-volt pulse of 30 ms duration was used to trigger the spin reversal process. A 20 $\mu$A current bias was applied to the Hall bars. The signals were filtered and amplified and continuously recorded using an analog to digital acquisition card. Experiments were carried out at a bath temperature of $T_0=0.4$ K in a $^3$He refrigerator with a 3D vector superconducting magnet capable of producing bipolar transverse magnetic fields of up to $H_x=8$ T and bipolar longitudinal magnetic fields of up to $H_z=0.7$ T. (See Fig. 1(b) for the definition of the coordinate system.)

Fig. 1(c) shows the evolution of the magnetization of the crystal during several experimental runs. The crystal was initially magnetized to negative saturation ($M/M_s=-1$) by applying a bias field of $-0.7$ T and a $4.5$~T transverse field. Then, $H_x$ is set to a particular value and $H_z$ is swept to a positive field. Because of the large magnetic anisotropy of Mn$_{12}$-ac, the magnetization of the crystal remains blocked (black curve).  At a set magnetic field ($(H_x,H_z)$, triangles), a trigger heat pulse is supplied to ignite the magnetization reversal. In Fig.~\ref{fig1}(c), the magnetization change is represented by the vertical arrows and the filled circles are the measured final magnetic state. After the reversal process, $H_z$ is increased. However, this does not change the Hall bar signals and thus the magnetization of the crystal is nearly constant as seen by the horizontal lines to the right of the filled circles. As opposed to previously reported results on magnetic deflation, it is important to note that the final state is not full saturated, i.e. $M/M_s<1$ for $H_z \lesssim 0.5$~T.  We will discuss this in detail below.  Finally, the sample is remagnetized to negative saturation to repeat the process for other $H_x$ and $H_z$ magnetic fields.

The Hall sensor array permits time and spatially resolved measurements of the spin reversal. Fig.~\ref{fig1}(d) shows an example of the resulting Hall sensor signals as a function of time, with the pulse applied at $t=0$. A peak in the Hall signal indicates an increase of the fringe field $B_y$, and therefore when the spin reversal front is at a particular sensor. As a peak first appears in the sensor closest to the heater (sensor 1) it is clear that the process initiates at the edge closest to the heater. It then moves away from the heater and throughout the crystal. The speed of the front can be calculated from the time difference between the maxima in the Hall sensor responses and the distance between Hall sensors. We also note that the final magnetic state of the crystal was measured $\sim3$~s after the event (filled circles in Fig.~\ref{fig1}(c)). This indicates that $\sim 3$~s after the heat pulse, the system has returned to a blocked state. A measurement with a thermometer near the sample shows that it takes $\sim 1$~s for the system to return to the bath temperature $T_0=0.4$~K.

\section{Results and Discussion}

\noindent {\bf A.  Speed of the Deflagration Front}
\vspace{0.1in}

\begin{figure}[t]
\includegraphics[width=\columnwidth]{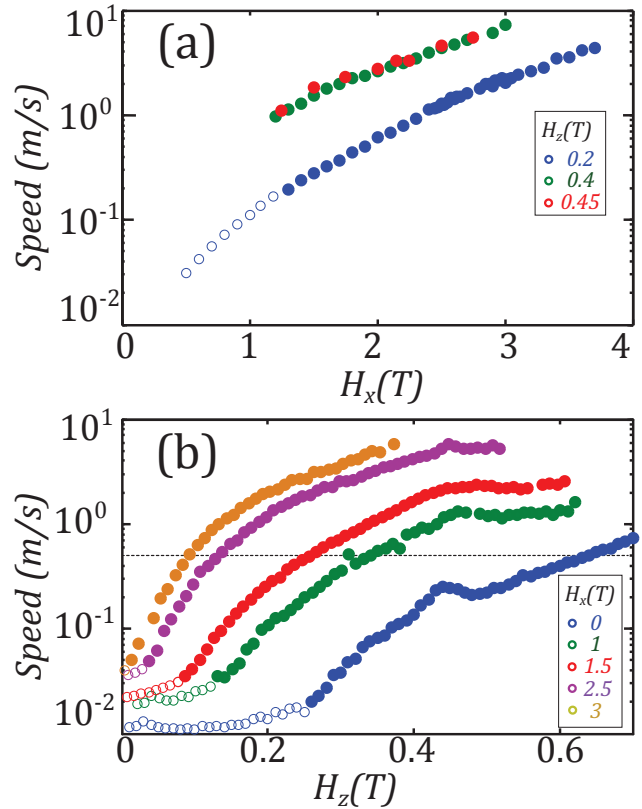}
\caption{\small{Speed of propagation of the deflagration front in Mn$_{12}$-ac:  (a) As a function of transverse magnetic field $H_x$ in several fixed longitudinal fields $H_z$. (b) As a function of longitudinal magnetic field $H_z$ in several fixed transverse fields $H_x$; open symbols are data taken in the thermal regime and closed symbols are taken in the deflagration regime.}}
\label{fig:speed}
\end{figure}

The speed of propagation of the deflagration front is shown in Fig.~\ref{fig:speed}(a) as a function of transverse field $H_x$ for various fixed longitudinal fields $H_z$; as noted earlier, varying the transverse field in fixed longitudinal field varies principally the activation energy, while not significantly affecting the energy released. Figure~\ref{fig:speed}(b) shows the propagation speed as a function of longitudinal field $H_z$ for various fixed transverse fields $H_x$; varying the longitudinal field changes both the activation energy and the Zeeman energy released. In both panels, the open symbols are data taken in the thermal regime and the closed symbols are in the deflagration regime; the change from open to closed symbols is thus the boundary between the two regimes.  Note that the longitudinal fields vary from zero to just above the first ($k=1$) resonant field of $0.45$~T,  thus considerably smaller than in previous experimental studies of magnetic deflagration.

As shown in Fig.~\ref{fig:speed}(b), once the deflagration regime is entered, the speed of the deflagration front varies nearly exponentially and then more slowly as the longitudinal field increases. The local maxima at $H_z\sim 0.45$~T are due to resonant quantum tunneling \cite{quantumdeflagration,McHugh2007}.  The speed of the front also increases rapidly at the larger values of transverse field.  This is seen more clearly in Fig.~\ref{fig:speed}(a), where the longitudinal field is fixed and the transverse field is varied. The data in the deflagration regime (solid symbols) show that the speed increases close to exponentially with increasing transverse field.

The rapid increase in the front speed with longitudinal field shown in Fig.~\ref{fig:speed}(b) is easily understood within the standard model for magnetic deflagration \cite{Garanin2007}. The deflagration process depends sensitively on the flame temperature, which depends on the energy released as $T_f\propto(\Delta E)^{1/4}$ \cite{Garanin2007}. The change in flame temperature is thus largest when $\Delta E$ is small, which is at small longitudinal fields. However, as the longitudinal field increases the change in flame temperature
$dT_f/d\Delta E$ decreases and this leads to a front speed that becomes a weaker function of the longitudinal field. Varying the transverse field at fixed longitudinal fields (Fig.~\ref{fig:speed}(a)) changes only the activation energy. In this case the temperature of the flame should be independent of the transverse field to first order.

An interesting aspect of the data in Fig.~\ref{fig:speed}(b) is that deflagration occurs with the same propagation speed for different applied field combinations ($H_x,H_z$).  This is clearly seen in Fig.~\ref{fig:speed}(b), where the dashed horizontal line denoting a propagation speed $0.5$ m/s occurs for different ($H_x,H_z$). This indicates that different combinations of activation and released energy lead to the same front speed. If the longitudinal field is small and the transverse field is large we have a small amount of energy released in spin-reversal but also a small activation barrier separating the spin-states. In such a case, spins can reverse rapidly because they require little energy to overcome the activation barrier. If we now consider the opposite case where the longitudinal field is larger and the transverse field is small, the spins cannot relax as fast as before because the activation energy is much larger. However, the larger longitudinal field means that the energy released is larger and this energy maintains the front speed. Thus one of the main differences between differently triggered deflagrations that have the same speed is the amount of energy released and, consequently, their flame temperatures. It is also expected that the width of the spin-reversing flame front $w_f$ will be different. However, because we are measuring the fringe field from a crystal that has a lateral size that is comparable to the width of the flame front, our experiment is not able to resolve  $w_f$ (see Appendix I).

\vspace{0.1in}
\noindent {\bf B.  Partial Spin Reversal}
\vspace{0.1in}

Previous experiments using high bias fields yielded a fully magnetized crystal as the final state.  High bias fields were found to be necessary to lower the activation barrier and enable the ignition of deflagration.  As noted earlier, the activation energy can be reduced by applying a transverse field. This enables ignition at small longitudinal bias, a range of field conditions not yet studied.  Our main finding in small longitudinal fields is that, while the magnetic deflagration encompasses the entire sample (i.e., the front does not stop in the sample interior), the final state is a homogeneous crystal that is only partially magnetized. Evidence for this is that there is a peak in each sensor, indicating a front passed over each sensor. Further, after the event the voltage level of each sensor can be compared to the voltage level measured in that same sensor when the sample is fully saturated. The fact that the ratio of these voltages is the same for each sensor indicates that the sample is uniformly magnetized. 

\begin{figure}[t]
\includegraphics[width=\columnwidth]{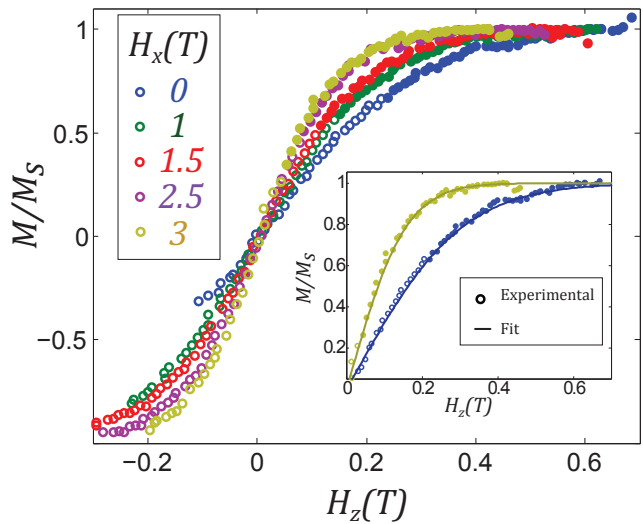}
\caption{\small{The main panel shows the final state of the magnetization of Mn$_{12}$-ac in different applied magnetic fields (both longitudinal and transverse). The open (closed) symbols are data taken in the thermal (deflagration) regime. The inset shows a fit of the data to a model described in the text. The solid lines are the fit.}}
\label{fig:finalstate}
\end{figure}

As described earlier in this paper, the final magnetization was recorded following deflagration, about $3$ seconds after the event (see Fig.~\ref{fig1}(c)) and data was acquired when the sample had returned to the bath temperature of $0.4$ K.  The results are plotted in Fig.~\ref{fig:finalstate} for five different values for the transverse field.  Note that data, shown as open symbols, are included for field conditions that do not ignite magnetic deflagration (e.g., at $H_x=0$ and $H_z<0.3$ T), corresponding to thermally driven relaxation \cite{subedi2013}.  The final magnetization shown in Fig.~\ref{fig:finalstate} is different for different transverse magnetic fields.  However, the {\em equilibrium final state} at the bath temperature would be nearly fully saturated for all cases we have studied, as $2g\mu_BSH_z/(k_BT)\gtrsim3$ for $H_z\gtrsim 0.04$~T, and thus the curves shown in different transverse fields would largely overlap. This indicates that the final states are not in equilibrium at the bath temperature. They are, in fact, magnetization states that are blocked once the sample temperature drops below the blocking temperature, a temperature set by the anisotropy barrier and an attempt frequency.

This can be understood as follows. During the magnetic deflagration the temperature of the propagating spin-reversal front increases to a flame temperature, $T_f$, above the blocking temperature (on a time scale of $\sim 1$~ms); as the magnetization equilibrates on a much shorter time scale at $T_f$, the magnetization at the flame front is in an equilibrium state determined by the longitudinal field and the temperature. Note that the transverse field does not significantly affect the equilibrium magnetization.  Following the deflagration, the sample cools down to the temperature of the bath. However, the magnetization does not have time to equilibrate at the bath temperature and blocks at a higher temperature (much larger than the bath temperature) determined by the competition between the decay time of the sample temperature, $T=(T_f-T_\mathrm{bath}) \exp\left[-\frac{t}{\tau}\right]$, and the spin relaxation rate $\Gamma=\Gamma_0\exp\left[\frac{U(H_x,H_z)}{k_BT}\right]$, which depends on \emph{both} longitudinal and transverse fields. Note that in the thermal regime at small bias fields (open symbols in Figs. \ref{fig:speed} and \ref{fig:finalstate}) magnetic relaxation is driven by the heat pulse, which also drives the sample above the blocking temperature. So in both cases (both thermal relaxation and magnetic deflagration regimes), the final state depends on the applied field under which the sample is cooled.

The final states shown in Fig.~\ref{fig:finalstate} vary with transverse field.  The equilibrium magnetic state depends principally on the longitudinal field, $H_z$, and the temperature of the bath, $T= 0.4$~K.  It is therefore clear that the final state at the bath temperature is a non-equilibrium magnetic state.  This can be traced to the fact the blocking temperature depends on the height of the barrier, $U(H_x,H_z)$ and since transverse fields reduce the barrier height, it also reduces the blocking temperature yielding different out-of-equilibrium final states.

The final state data can be fit by determining the blocking temperature at a given field configuration and considering the time scale that sets the rate of change of the sample temperature, $\tau$. We calculate the relaxation rate $\Gamma(H_z,H_x,T)$ at a given field, with $U(H_x,H_z)$ (Eq.~\ref{barrier}), to find the temperature, $T_b$ at which the blocking condition is satisfied:
\beq
\Gamma(H_x,H_z,T_b)=1/ \tau.
\eeq
We further assume that the magnetization is in equilibrium at the blocking temperature and does not evolve further as the  sample temperature decreases to the bath temperature. We plot the equilibrium magnetization at the blocking temperature in the inset of Fig.\ \ref{fig:finalstate} with a  decay time of about 1 second. The agreement between calculated curves and experiments is excellent.

\vspace{0.1in}
\noindent {\bf C.  Proposed Model}
\vspace{0.1in}

Magnetic deflagration in molecular magnets has been investigated theoretically in Refs.~\cite{Garanin2007,modestovprl}.  We extend the theory by including the dependence of the activation barrier and the energy released on the transverse field, $U(H_x, H_z)$, $\Delta E(H_x, H_z)$. We also account for the fact that the final state and thus the magnetization at the deflagration front is not a fully saturated magnetic state, but instead is one in which the spins in the front are only partially reversed (i.e., $M/M_s <1$). This extended model of magnetic deflagration is presented in Appendix II.

\begin{figure}[t]
\includegraphics[width=\columnwidth]{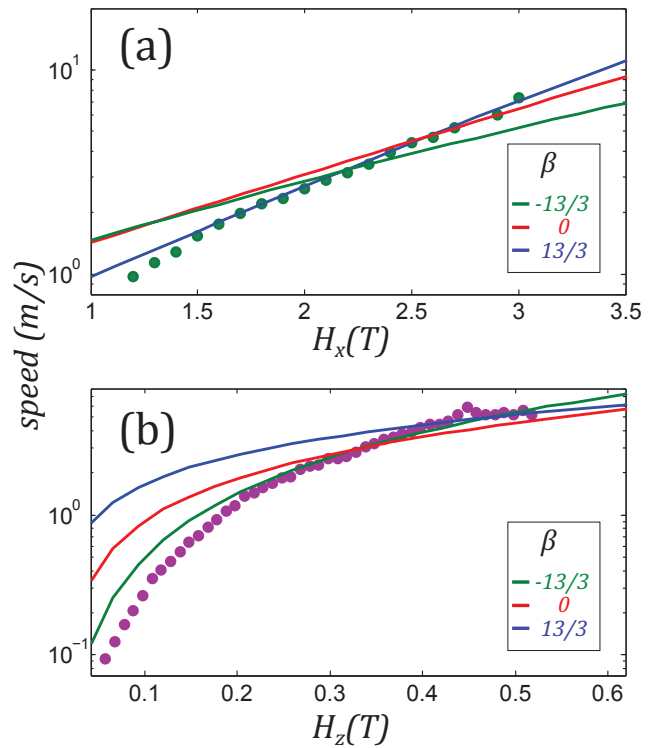}
\caption{\small{Speed of propagation of the deflagration front in Mn$_{12}$-ac as a function of the applied field and fits to deflagration theory for three different values of the exponent $\beta$ in the thermal diffusivity, $\kappa \propto T^{-\beta}$. (a) The transverse field dependence of the  front speed at a fixed longitudinal field of $H_z =0.4$ T. (b) The longitudinal field dependence of the front speed at a fixed transverse field of $H_x =2.5$ T. }}
\label{fig:model}
\end{figure}

Interestingly, our model fits the speed versus transverse field data (Fig.~\ref{fig:model}(a)) with a thermal diffusivity, $\kappa$,  that decreases with increasing temperature, $\kappa\propto T^{-\beta}$ with $\beta=13/3$, very well. This exponent is expected based on the temperature dependence of the thermal conductivity and heat capacity at low temperature \cite{Garanin2007,modestovprl}. However, the theory does not fit the speed versus longitudinal field data (Fig.~\ref{fig:model}(b)), with this same temperature dependence of the diffusivity. We obtain better fits when considering the thermal diffusivity to increase with the temperature. Fig.~\ref{fig:model}(b) shows three fits to the data using different temperature dependences for the thermal diffusivity, $\beta=13/3$, $0$ and $-13/3$. This discrepancy between magnetic deflagration theory and experiment has already been noted in previous experiments in which the speed of the front was measured as a function of longitudinal fields at much larger fields \cite{McHugh2009}.

\section{Conclusion}

To summarize, this paper reports a systematic study of the effect of transverse magnetic field on magnetic deflagration in Mn$_{12}$-ac.  Agreement with theory is found for the speed of propagation of the deflagration front as a function of transverse field in fixed longitudinal field. However, and as reported in earlier studies \cite{McHugh2009}, the same theory does not fit the front speed as a function of longitudinal magnetic field in fixed transverse field, suggesting extensions of the model may be necessary. 

A particularly interestingly result is that experiments conducted in large transverse fields and small longitudinal fields show clear evidence for partial spin-reversal in magnetic deflagration.  As the flame temperature is higher than the bath temperature and the blocking temperature, our measurements and analysis demonstrate that the magnetization at the deflagration front is also not fully reversed.  This suggests that even in experiments that result in a fully magnetized crystal, the magnetization at the deflagration front may not be fully reversed, a fact that needs to be considered for a full understanding of the process.  Moreover, and perhaps more significantly, an unsaturated magnetization at the flame front is a necessary condition for observing internal dynamics of the front, such as oscillations \cite{Modesto2011} and thermal instabilities.  We expect partial spin reversal to be quite general to deflagration in magnetic systems, opening the possibility of observing internal front dynamics.

\section*{Acknowledgements}
S.V. acknowledges financial support from the Ministerio de Educación, Cultura y Deporte de España.
F.M. acknowledges support from a Marie Curie IOF 253214. MPS acknowledges support by NSF-DMR-0451605 and ARO W911NF- 13-1-0125. Research at NYU was supported in part by NSF-DMR-1006575, DMR-1309202 and NYU.

\section{Appendix I: Limitations in Magnetic Measurements of the Front Width of a Magnetic Deflagration}
\label{Appendix2}
The \emph{flame width} in magnetic deflagration depends on the thermal diffusivity, $\kappa$, and the relaxation rate, $\Gamma$, at the temperature of the flame, $w_f=\sqrt{\kappa_f/\Gamma_f}$. Thus a variation in the flame temperature directly affects the flame width. So if $\kappa$ were a constant (i.e. independent of temperature), the flame width would narrow with increasing flame temperature. Then same-speed deflagration fronts for different magnetic fields would have different flame widths. However, our method of measuring the fringe field is limited in spatial resolution by the lateral dimensions of the crystal. This is because the fringe field even from an infinitely sharp front at the position of our Hall sensor would have a width of order of the crystal lateral scale due to the spread of dipole fields. Thus, as long as the front width is smaller than the crystal width our measurement method is relatively insensitive to the width of the front.

\begin{figure}[h]
\includegraphics[width=\columnwidth]{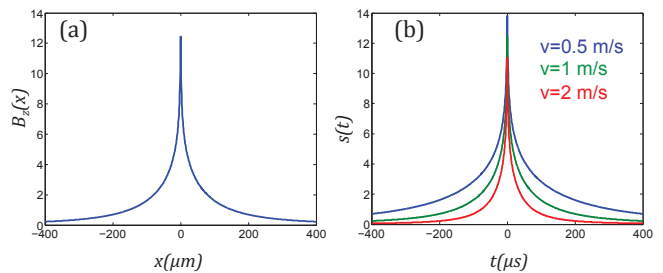}
\caption{\small{(a) The field distribution $B_z(x)$ for a crystal with a infinitely sharp deflagration front centered at $x=0$. We take a crystal width of $\omega=200\:\mu$m.  (b) Hall sensor signal $s$ versus time for speeds of $v=0.5$, $v=1$, and $2$ m/s. The horizontal axes in (a) and (b) are related by $x=vt$.}}
\label{fig1Annex}
\end{figure}

Consider an infinitely sharp front propagating at a constant speed $v$ in a crystal with a lateral dimension, $w$.
The fringe fields in the $z$ direction are given by
\beq
B_z(x)=M\log\left(\frac{x^2+w^2}{x^2}\right).
\eeq
The measured signal for a moving front in the Hall sensors is $s(t)=B_z(vt)$.
For example, Fig.\ \ref{fig1Annex} plots both $B_z(x)$ and $s(t)$  in a crystal with $\omega=200\:\mu$m for speeds of $v=0.5$, $1$, and $2$ m/s.

There are two more sources of broadening of the peak measured with Hall probes in addition to the spread in $B_z$ and the speed of the front $v$: i) the effective area of the Hall probes (about 20 $\mu$m) and ii) the width of the flame $w_f$. In the first case we just convolute a single pulse function with the length of the Hall probe active area \cite{Lui1998} with $B_z(x)$. In the second case we should compute the function $B_z$ for a front that is not infinitely sharp. Alternatively, we could consider a Gaussian (or a Lorentzian) shaped peak, $f_p(x)$, with a width given by the flame width and convolute it with $B_z(x)$.s

The convolution of two peaks with widths $w_1$ and $w_2$ is approximately $\sqrt{w_1^2+w_2^2}$. (This an equality in the case of Gaussians.)

In summary, we could convert measured peaks with our Hall probes from $s(t)$ to $B_z(x/v)$. (Assuming Gaussian peaks we would simply multiply the width in time $\Delta t$ by the speed $v$ to get the width in space.) However, the value we obtain for the width includes the width of the fringe field distribution, the finite spatial resolution of the sensors, and the width of the flame in the following approximate form $\sqrt{w^2+w_f^2+w_{Hp}^2}$.

Notice that when the width of the flame becomes smaller than the crystal dimension the fringe field width dominates the width of the peaks we measure. Hence we are not able to accurately determine a flame width that is smaller than the crystal width. This is likely the reason we observe a width ($v\Delta t$) that is nearly independent of the bias field, i.e. the flame width is actually changing but our measurement is dominated by the spread in fringe fields.

\section{Appendix II: Speed of the Deflagration Front as a Function of $H_x$,  $H_z$}

We consider the following system of equations for the phonon energy $\mathcal{E}$ and the number of spins in the metastable state $n$ that describe the deflagration process \cite{Garanin2007,modestovprl,Modesto2013}:
\beq
\begin{array}{rcl}
\ds\frac{\p \mathcal{E}}{\p t}&=&\ds\nabla\cdot\kappa\nabla\mathcal{E}-\Delta E \frac{\p n}{\p t}\\\\
\ds\frac{\p n}{\p t}&=&\ds-\Gamma(n-n_{\text{eq}}).
\end{array}
\label{eqs}
\eeq
Here $\Delta E$  is the Zeeman energy, $\kappa$ is the thermal diffusivity, which depends on the temperature as $\kappa\propto T^{-\beta}$, $\Gamma= \Gamma_0 \exp[-U/T]\left[1+\exp(-\Delta E/T)\right]$ is the relaxation rates, $U$ the activation barrier,  $\Gamma_0$ is a constant attempt rate, and $n_{\text{eq}}=1/(1+e^{\frac{\Delta E}{T}})$.

The relation between the temperature and the phonon energy is
\beq
\mathcal{E}=\frac{A\Theta_D}{\alpha+1}\left(\frac{T}{\Theta_D}\right)^{\alpha+1},
\eeq
where $\Theta_D$ is the Debye temperature. The coefficient $A$ is a constant, $A = 13\pi^4/5 \approx 234$ and the exponent $\alpha$ is taken to be $3$.

The energy barrier, $U$, and the Zeeman energy, $\Delta E$, as a function of the applied fields $H_x$ and $H_z$ are determined as follows. We use a classical expression for the energy landscape, which neglects resonant quantum tunneling of the magnetization, but otherwise should accurately capture the dependence of the activation energy on the applied field. The energy is normalized to $DS^2$ and the fields are normalized to the anisotropy field ($H_A=2DS/(g\mu_B)$), see \cite{subedi2013}). The energy as a function of field and the angle of magnetization relative to the easy axis $\theta$ is:
\beq
E(h_x,h_z,\theta)=\frac{1}{2}\sin^2\theta-h_z\cos\theta-h_x\sin\theta.
\label{barrier}
\eeq
\\
From the minima and maxima of $E$ as a function of $\theta$ the activation energy and energy barrier are determined.

The temperature of the flame is found by energy conservation $\mathcal{E}_0+\Delta E n_0=\mathcal{E}_f+\Delta E n_f$ that gives a simple equation if $n_0=1$ and $n_f=0$ (i.e. if the initial state is all spins in the metastable state and the final state is fully saturated, all the spins reversed)
\beq
T_f=\Theta_D\left(\frac{4\Delta E}{A\Theta_D}\right)^{1/4},
\eeq
but for a partially reversed final state one has to solve the equation
\beqa
\frac{A\Theta_D}{4}\left(\frac{T_0}{\Theta_D}\right)^{4}+\Delta En_0=\frac{A\Theta_D}{4}\left(\frac{T_f}{\Theta_D}\right)^{4} \\ \nonumber +\Delta E\left(1-\frac{1}{1+e^{\frac{\Delta E}{T_f}}}\right).
\eeqa
Note that it is the final state at the flame temperature, $n_{\text{eq}}=n_f$, that is relevant.

We normalize equations \ref{eqs} following Ref.~\cite{Garanin2007}
\beq
\tilde{\mathcal{E}}=\frac{\mathcal{E}}{\Delta E},\, \, \, \tau=t\Gamma_f, \,\, \tilde{\mb{r}}=\frac{\mb{r}}{l_d}\,\, \tilde{T}=\frac{T}{T_f}, \, \, \tilde{\kappa}=\frac{\kappa}{\kappa_f},
\eeq
where $T_f$ and $\kappa_f$ are the temperature of the flame and the thermal diffusivity at $T_f$, and $l_d=\sqrt{\kappa_f/\gamma_f}$. The resulting Eqs.~\ref{eqs} read:
\beq
\begin{array}{rcl}
\ds\frac{\p \tilde{\mathcal{E}}}{\p \tau}&=&\ds\tilde{\nabla}\cdot\tilde{\kappa}\tilde{\nabla}\tilde{\mathcal{E}}- \frac{\p \tilde{n}}{\p \tau}\\\\
\ds\frac{\p \tilde{n}}{\p \tau}&=&\ds-\tilde{\Gamma}\left(\tilde{n}-\frac{1}{1+e^{\frac{\Delta E}{T}}}\right),
\end{array}
\label{eqs2}
\eeq
We consider a moving flat deflagration front as a solution of Eqs.\ \ref{eqs2} that depends on the combined timelike argument $u=\tau-\tilde{x}/\tilde{v}$. So Eqs. \ref{eqs2} take the form
\beq
\frac{\p \tilde{\mathcal{E}}}{\p u} = \frac{1}{\tilde{v}^2}\frac{d}{du}\tilde{\kappa}\frac{d\tilde{\mathcal{E}}}{du}-\frac{d\tilde{n}}{du}, \, \,\,\,\,\, \frac{d\tilde{n}}{du}=\tilde{\Gamma}\tilde{n}
\label{norm}
\eeq
The real deflagration speed $v$ is given by
\beq
v=\tilde{v}\sqrt{\kappa_f\gamma_f}.
\eeq
The first of Eqs.\ \ref{norm} can be integrated to give:
\beq
\frac{\p \tilde{\mathcal{E}}}{\p u} = \frac{\tilde{v}^2}{\tilde{k}}\left(\tilde{\mathcal{E}}+\tilde{n}-1\right), \,\,\,\,\, \frac{d\tilde{n}}{du}=\tilde{\Gamma}\tilde{n}.
\eeq
Numerical solution requires imposing proper boundary conditions and solving for $\tilde{v}$. A physical speed $\tilde{v}$ is a solution with a front to the right of $\tilde{n}=1$ with $\tilde{\mathcal{E}}$ having an asymptotic form that reflects the temperature changing from $T$ to $T_f$.

The parameters used to fit data in Fig.~\ref{fig:model} are $\kappa=3\times10^{-2} T^{-13/3}$, $1.5\times10^{-5}$ and $3\times10^{-9} T^{13/3}$ m$^2$/s.

\end{document}